\input harvmac
\input graphicx
\input color

\def\Title#1#2{\rightline{#1}\ifx\answ\bigans\nopagenumbers\pageno0\vskip1in
\else\pageno1\vskip.8in\fi \centerline{\titlefont #2}\vskip .5in}

%
%
\ifx\includegraphics\UnDeFiNeD\message{(NO graphicx.tex, FIGURES WILL BE IGNORED)}
\def\figin#1{\vskip2in}
\else\message{(FIGURES WILL BE INCLUDED)}\def\figin#1{#1}
\fi
\def\Fig#1{Fig.~\the\figno\xdef#1{Fig.~\the\figno}\global\advance\figno
 by1}
%
%
%
%

\font\ticp=cmcsc10

\def \purge#1 {\textcolor{magenta}{#1}}
\def \new#1 {\textcolor{blue}{#1}}
\def\comment#1{}

\def\\{\cr}
\def\text#1{{\rm #1}}
\def\frac#1#2{{#1\over#2}}

\def\subsubsec#1{\noindent{\undertext {#1}}}
\def\undertext#1{$\underline{\smash{\hbox{#1}}}$}
\def\calo{{\cal O}}

\def\vac{|0\rangle}
\def\vacb{\langle0|}

\def\calh{{\cal H}}
\def\hext{{{\cal H}_{\rm ext}}}

\def\hreg{{{\cal H}_{\rm reg}}}
\def\hM{{\cal H}_{\rm M}}

\def\roughly#1{\mathrel{\raise.3ex\hbox{$#1$\kern-.75em\lower1ex\hbox{$\sim$}}}}
\font\bbbi=msbm10 
\def\mathbb#1{\hbox{\bbbi #1}}

\def\AA{{\cal A}_A}

 \def\tX{{\tilde X}}

\def\mthsu{\mathsurround=0pt  }
\def\leftrightarrowfill{$\mthsu \mathord\leftarrow\mkern-6mu\cleaders
  \hbox{$\mkern-2mu \mathord- \mkern-2mu$}\hfill
  \mkern-6mu\mathord\rightarrow$}
\def\overleftrightarrow#1{\vbox{\ialign{##\crcr\leftrightarrowfill\crcr\noalign{\kern-1pt\nointerlineskip}$\hfil\displaystyle{#1}\hfil$\crcr}}}
\overfullrule=0pt

%
%
\lref\tHoo{
  G.~'t Hooft,
  ``Dimensional reduction in quantum gravity,''
  [gr-qc/9310026].
}
\lref\Susstrans{
  L.~Susskind,
  ``The Transfer of Entanglement: The Case for Firewalls,''
[arXiv:1210.2098 [hep-th]].
}
\lref\vanR{
  M.~Van Raamsdonk,
  ``Building up spacetime with quantum entanglement,''
Gen.\ Rel.\ Grav.\  {\bf 42}, 2323 (2010), [Int.\ J.\ Mod.\ Phys.\ D {\bf 19}, 2429 (2010)].
[arXiv:1005.3035 [hep-th]].
}
\lref\Susstrouble{
  L.~Susskind,
  ``Trouble for remnants,''
[hep-th/9501106].
}
\lref\Sussholo{
  L.~Susskind,
  ``The World as a hologram,''
J.\ Math.\ Phys.\  {\bf 36}, 6377 (1995).
[hep-th/9409089].
}
\lref\SusskindIF{
  L.~Susskind, L.~Thorlacius and J.~Uglum,
 ``The Stretched horizon and black hole complementarity,''
Phys.\ Rev.\ D {\bf 48}, 3743 (1993).
[hep-th/9306069].
}
\lref\AMPS{
  A.~Almheiri, D.~Marolf, J.~Polchinski and J.~Sully,
  ``Black Holes: Complementarity or Firewalls?,''
  JHEP {\bf 1302}, 062 (2013).
  [arXiv:1207.3123 [hep-th]].
}
\lref\HaPr{
  P.~Hayden, J.~Preskill,
  ``Black holes as mirrors: Quantum information in random subsystems,''
JHEP {\bf 0709}, 120 (2007).
[arXiv:0708.4025 [hep-th]].
}
\lref\MaSu{
  J.~Maldacena and L.~Susskind,
  ``Cool horizons for entangled black holes,''
Fortsch.\ Phys.\  {\bf 61}, 781 (2013).
[arXiv:1306.0533 [hep-th]].
}
\lref\Mathurrev{
  S.~D.~Mathur,
  ``Fuzzballs and the information paradox: A Summary and conjectures,''
[arXiv:0810.4525 [hep-th]].
}
\lref\HST{
  T.~Banks,
  ``Lectures on Holographic Space Time,''
[arXiv:1311.0755 [hep-th]].
}
\lref\GiShtwo{
  S.~B.~Giddings and Y.~Shi,
  ``Effective field theory models for nonviolent information transfer from black holes,''
[arXiv:1310.5700 [hep-th]], Phys.\ Rev.\ D (in press).
}
\lref\NVNLT{
  S.~B.~Giddings,
  ``Modulated Hawking radiation and a nonviolent channel for information release,''
[arXiv:1401.5804 [hep-th]].
}
\lref\QBHB{
  S.~B.~Giddings,
  ``Quantization in black hole backgrounds,''
Phys.\ Rev.\ D {\bf 76}, 064027 (2007).
[hep-th/0703116 [HEP-TH]].
}
\lref\BHMR{
  S.~B.~Giddings,
  ``Black holes and massive remnants,''
Phys.\ Rev.\ D {\bf 46}, 1347 (1992).
[hep-th/9203059].
}
\lref\BHIUN{
  S.~B.~Giddings,
  ``Black hole information, unitarity, and nonlocality,''
Phys.\ Rev.\ D {\bf 74}, 106005 (2006).
[hep-th/0605196].
}
\lref\BHQIUE{
  S.~B.~Giddings,
  ``Black holes, quantum information, and unitary evolution,''
  Phys.\ Rev.\ D {\bf 85}, 124063 (2012).
[arXiv:1201.1037 [hep-th]].
}
\lref\SGmodels{
  S.~B.~Giddings,
   ``Models for unitary black hole disintegration,''  Phys.\ Rev.\ D {\bf 85}, 044038 (2012)
[arXiv:1108.2015 [hep-th]].
}
\lref\NLvC{
  S.~B.~Giddings,
  ``Nonlocality versus complementarity: A Conservative approach to the information problem,''
Class.\ Quant.\ Grav.\  {\bf 28}, 025002 (2011).
[arXiv:0911.3395 [hep-th]].
}
\lref\NLEFTone{
  S.~B.~Giddings,
  ``Nonviolent information transfer from black holes: a field theory parameterization,''
Phys.\ Rev.\ D {\bf 88}, 024018 (2013).
[arXiv:1302.2613 [hep-th]].
}
\lref\NVNL{
  S.~B.~Giddings,
  ``Nonviolent nonlocality,''
  Phys.\ Rev.\ D {\bf 88},  064023 (2013).
[arXiv:1211.7070 [hep-th]].
}
\lref\BHSM{
  S.~B.~Giddings,
  ``Statistical physics of black holes as quantum-mechanical systems,''
Phys.\ Rev.\ D {\bf 88}, 104013 (2013).
[arXiv:1308.3488 [hep-th]].
}
\lref\GiShone{
  S.~B.~Giddings and Y.~Shi,
  ``Quantum information transfer and models for black hole mechanics,''
Phys.\ Rev.\ D {\bf 87}, 064031 (2013).
[arXiv:1205.4732 [hep-th]].
}
\lref\Pres{
  J.~Preskill,
  ``Do black holes destroy information?,''
  in proceedings of Black holes, membranes, wormholes and superstrings, Houston 1992
[hep-th/9209058].
}
\lref\WABHIP{
  S.~B.~Giddings,
  ``Why aren't black holes infinitely produced?,''
Phys.\ Rev.\ D {\bf 51}, 6860-6869 (1995).
[hep-th/9412159].
}
\lref\Hawk{
  S.~W.~Hawking,
  ``Particle Creation By Black Holes,''
  Commun.\ Math.\ Phys.\  {\bf 43}, 199 (1975)
  [Erratum-ibid.\  {\bf 46}, 206 (1976)].
}
\lref\LPSTU{
  D.~A.~Lowe, J.~Polchinski, L.~Susskind, L.~Thorlacius and J.~Uglum,
  ``Black hole complementarity versus locality,''
Phys.\ Rev.\ D {\bf 52}, 6997 (1995).
[hep-th/9506138].
}
\lref\PaRa{
  K.~Papadodimas and S.~Raju,
  ``An Infalling Observer in AdS/CFT,''
[arXiv:1211.6767 [hep-th]].
}
\lref\VV{
  E.~Verlinde and H.~Verlinde,
  ``Black Hole Entanglement and Quantum Error Correction,''
[arXiv:1211.6913 [hep-th]].
}
\lref\Braunstein{
  S.~L.~Braunstein, S.~Pirandola and K.~\.Zyczkowski,
  ``Entangled black holes as ciphers of hidden information,''
Physical Review Letters 110, {\bf 101301} (2013).
[arXiv:0907.1190 [quant-ph]].
}
\lref\DvaliAA{
  G.~Dvali and C.~Gomez,
  ``Black Hole's Quantum N-Portrait,''
Fortsch.\ Phys.\  {\bf 61}, 742 (2013).
[arXiv:1112.3359 [hep-th]].
}
\lref\DvaliEJA{
  G.~Dvali and C.~Gomez,
  ``Quantum Compositeness of Gravity: Black Holes, AdS and Inflation,''
[arXiv:1312.4795 [hep-th]].
}
\lref\miningrefs{
  W.~G.~Unruh and R.~M.~Wald
  ``How to mine energy from a black hole,''
Gen.\ Relat.\ Gravit.\ {\bf 15}, 195 (1983)\semi
  A.~E.~Lawrence and E.~J.~Martinec,
  ``Black hole evaporation along macroscopic strings,''
Phys.\ Rev.\ D {\bf 50}, 2680 (1994)
[hep-th/9312127]\semi
  V.~P.~Frolov and D.~Fursaev,
  ``Mining energy from a black hole by strings,''
Phys.\ Rev.\ D {\bf 63}, 124010 (2001)
[hep-th/0012260]\semi
  V.~P.~Frolov,
 ``Cosmic strings and energy mining from black holes,''
Int.\ J.\ Mod.\ Phys.\ A {\bf 17}, 2673 (2002).
}
\lref\DeBo{
  S.~Deser and D.~Boulware,
  ``Stress-Tensor Commutators and Schwinger Terms,''
J.\ Math.\ Phys.\  {\bf 8}, 1468 (1967).
}
\lref\MaPl{
  S.~D.~Mathur and C.~J.~Plumberg,
 ``Correlations in Hawking radiation and the infall problem,''
JHEP {\bf 1109}, 093 (2011).
[arXiv:1101.4899 [hep-th]].
}
\lref\BLR{
  A.~E.~Broderick, A.~Loeb and R.~Narayan,
  ``The Event Horizon of Sagittarius A*,''
Astrophys.\ J.\  {\bf 701}, 1357 (2009).
[arXiv:0903.1105 [astro-ph.HE]].
}
\lref\BJLP{
  A.~E.~Broderick {\it et al.}  [Perimeter Institute for Theoretical Physics Collaboration],
  ``Testing the No-Hair Theorem with Event Horizon Telescope Observations of Sagittarius A*,''
Astrophys.\ J.\  {\bf 784}, 7 (2014).
[arXiv:1311.5564 [astro-ph.HE]].
}
\lref\Fishetal{
  V.~Fish, W.~Alef, J.~Anderson, K.~Asada, A.~Baudry, A.~Broderick, C.~Carilli and F.~Colomer {\it et al.},
  ``High-Angular-Resolution and High-Sensitivity Science Enabled by Beamformed ALMA,''
[arXiv:1309.3519 [astro-ph.IM]].
}
\lref\AMPSS{
  A.~Almheiri, D.~Marolf, J.~Polchinski, D.~Stanford and J.~Sully,
  ``An Apologia for Firewalls,''
JHEP {\bf 1309}, 018 (2013).
[arXiv:1304.6483 [hep-th]].
}
\lref\JoPs{
  T.~Johannsen and D.~Psaltis,
  ``Testing the No-Hair Theorem with Observations in the Electromagnetic Spectrum: II. Black-Hole Images,''
Astrophys.\ J.\  {\bf 718}, 446 (2010).
[arXiv:1005.1931 [astro-ph.HE]].
}
\lref\Bardeen{J. Bardeen,  ``Timelike and null geodesics in the Kerr metric," in {\it Black Holes}, eds. DeWitt \& DeWitt (1973). }
\lref\Falcke{
  H.~Falcke, F.~Melia and E.~Agol,
  ``Viewing the shadow of the black hole at the galactic center,''
  Astrophys.\ J.\ {\bf 528} L13 (2000).
[astro-ph/9912263].
}
\lref\MTW{
  C.~W.~Misner, K.~S.~Thorne and J.~A.~Wheeler,
  ``Gravitation,'' (see p. 879)
San Francisco 1973.
}
\lref\BaFr{
  C.~Bambi and K.~Freese,
  ``Apparent shape of super-spinning black holes,''
Phys.\ Rev.\ D {\bf 79}, 043002 (2009).
[arXiv:0812.1328 [astro-ph]].
}
\lref\BaYo{
  C.~Bambi and N.~Yoshida,
  ``Shape and position of the shadow in the $\delta = 2$ Tomimatsu-Sato space-time,''
Class.\ Quant.\ Grav.\  {\bf 27}, 205006 (2010).
[arXiv:1004.3149 [gr-qc]].
}
\lref\SBGlett{S.~B.~Giddings, ``A final note on the existence of event horizons," (letter), Phys.\ Tod.\ {\bf 67}, no.\ 6, p11 (2014).}

\Title{
\vbox{\baselineskip12pt  
}}
{\vbox{\centerline{Possible observational windows for quantum} \centerline{effects from black holes}
}}

\centerline{{\ticp 
Steven B. Giddings\footnote{$^\ast$}{Email address: giddings@physics.ucsb.edu}
} }
\centerline{\sl Department of Physics}
\centerline{\sl University of California}
\centerline{\sl Santa Barbara, CA 93106}
\vskip.10in
\centerline{\bf Abstract}

Quantum information transfer necessary to reconcile black hole evaporation with quantum mechanics, while approximately preserving regular near-horizon geometry, can be simply parameterized in terms of couplings of the black hole internal state to  quantum fields of the black hole atmosphere. The necessity of transferring sufficient information for unitarization sets the strengths of these couplings. Such couplings via the stress tensor offer apparently significant advantages, and behave like quantum fluctuations of the effective metric near the horizon.   At the requisite strength, these fluctuations, while  soft (low energy/momentum), have significant magnitude, and so can deflect near-horizon geodesics that span distances of order the black hole radius. 
Thus, the presence of such couplings can result in effects that could be detected or constrained by observation: disruption of near-horizon accretion flows, scintillation of light passing close to the black hole, and alteration of gravitational wave emission from inspirals.  These effects could in particular distort features of Sgr A* expected to be observed, {\it e.g.}, by the Event Horizon Telescope, 
such as the black hole shadow and photon ring.

\vskip.3in
\Date{}

\newsec{Introduction}

The unitarity crisis associated with black hole formation and evaporation represents a profound clash between the foundational principles of modern physics.  These are the principles of relativity, the principles of quantum mechanics, and the principle of locality.  Our failure to reconcile these principles in the black hole context indicates one or more of them requires modification.  Consideration of the relative difficulties of different modifications strongly suggests that the principles of quantum mechanics should be retained, but at the price of the usual local quantum field theory (LQFT) notion of locality.

If field theory locality -- conventionally formulated as commutativity of observables outside the light cone, forbidding superluminal signaling -- must be modified, a key question is {\it how} it is modified.  Generic modifications of locality are expected to lead to disaster, by producing acausality, and consequent inconsistency.  The needed modification is expected to be quite special, if the resulting theory is to consistently reproduce local quantum field theory as a good approximation in a correspondence limit.

Different kinds of modification of locality have been considered.  One idea is to modify the notion of {\it localization}, and consider the possibility that quantum information can be localized in a very different fashion than in LQFT, depending for example on the nature of the observer.  Black hole complementarity\refs{\tHoo\Sussholo-\SusskindIF}\ exemplifies  such modification of localization.  More recent efforts to rescue aspects of complementarity from problems pointed out in \AMPS\ (see also \NLvC), {\it e.g.} via proposed relations between entanglement and geometry\refs{\vanR,\MaSu}, also are in this category.

A second way to modify LQFT locality is to retain a notion of localization, but to allow propagation or transfer of quantum information that can be, in certain circumstances, outside the light cone of the semiclassical geometry associated with a configuration.  Ultimately this probably arises from breakdown of the fundamental notion of geometry, and if a geometrical description does fail, an obvious question is how to describe that localization.  

A general approach to this, in the framework of quantum mechanics, is to retain a notion of localization as arising from a subsystem structure of the physical system.  This naturally generalizes the localization of LQFT, and suggests the possibility of a quantum generalization\BHQIUE\ of classical manifold structure.  While questions regarding the role of such a subsystem structure remain in the gravitational context, assuming that it gives a good approximate description appears useful and has been widely pursued.  

To give a quantum-mechanical description of physics, we also assume that there is a well-defined notion of unitary evolution.  
If classical spacetime approximately emerges from such a coarser-grained localization, the evolution may well not respect the corresponding approximate classical causal structure.  From the perspective of the spacetime geometry, this may be described by the  more fundamental hamiltonian transferring information outside the usual light cone.  Different approaches to the unitarity crisis  have explored this possibility, beginning with \BHMR.  In particular, such transfer is implicit in the ``firewall" scenario\AMPS\ and in an apparently more conservative variant of the ``fuzzball\refs{\Mathurrev}" scenarios.  

Taking this perspective, there are two approaches to describing  evolution of a quantum system like a black hole.  
One is to attempt to do so via such a  fundamental description, {\it e.g.} in terms of some quantum generalization of geometry\refs{\BHQIUE,\HST}.  This requires further development of the fundamental principles of the theory.  But, if LQFT evolution in a semiclassical geometry provides a good {\it approximate} description of the dynamics, for many purposes, a second approach presents itself:  parameterize the deviations from LQFT in an effective description within the framework of LQFT\refs{\NLEFTone\GiShtwo-\NVNLT}.  One of course expects such deviations to become significant for black holes followed over sufficiently long timescales (as well as near the center of a black hole) but the usual physics of LQFT may furnish a good approximation, up to ``small" corrections, for sufficiently small regions, away from the black hole core.

This paper will adopt the latter, effective, approach.  In particular, one may parameterize the information transfer necessary to unitarize black hole evaporation in terms of couplings of the internal quantum states of the black hole to the fields in the black hole atmosphere.  Generic such couplings also lead to increased energy flux\refs{\SGmodels,\BHQIUE,\NVNL,\NLEFTone,\GiShtwo} from the black hole, beyond that of Hawking\refs{\Hawk}: the information transfer is accompanied by energy transfer.  This produces a departure from expectations of black hole thermodynamics\refs{\BHSM}.  While this departure may possibly not be fundamentally inconsistent, an attempt to avoid it suggests consideration of special couplings to the fields, through the stress tensor\refs{\NVNL,\NLEFTone}, which can transfer information by modulating the pre-existing Hawking flux\refs{\NVNLT}.

The current work explores the effects of these stress-tensor couplings, which from the exterior perspective behave like extra quantum metric fluctuations in the atmosphere region near the horizon. The next section overviews the effective approach that has just been described, and discusses characterization of the necessary information transfer from the black hole.  Section three focusses specifically on stress-tensor couplings, and argues that information transfer sufficient to unitarize black hole evaporation via this mechanism requires  couplings of order unity.  This means that fluctuations in the effective metric become substantial, though they can be {\it soft} (low momentum).  As a result, there can be significant corrections to propagation of matter and light in the black hole atmosphere.  The size of these effects is estimated in the geodesic approximation.  Possible observational signatures are described,  including disruption of accretion disk dynamics near the horizon,  fluctuating contributions to lensing, or ``scintillation," of light, and distortion of gravity-wave signatures from inspirals.   Such effects may already be constrained by observational windows on near-horizon accretion flows, and could for example alter expected features of electromagnetic images of black holes, such as the ``black hole shadow" and ``photon ring." Near-term observations, such as via the Event Horizon Telescope, are expected to be sensitive to these features for Sgr A*, the black hole at the center of our galaxy, opening a new window for  observation.  Possible distortion of the photon ring, which is sensitive to the metric near the would-be photon orbit -- at 3/2 the Schwarzschild radius in the small angular momentum limit --  is a particularly promising target as a probe of the close-in geometry.  
Section four closes with some further perspective and with some comparison to other proposed scenarios for unitarization, and to their potential to produce observational effects.

\newsec{Effective description of information transfer}

\subsec{Subsystems and interactions}

We take as a starting point the assumption that information can be at least approximately localized in subsystems, which have an approximate correspondence with localized regions of spacetime. Specifically, for a black hole (BH) of mass $M$ and its surroundings, a coarsest form of such a localization arises if the states are approximately described by elements of
 a tensor product,
\eqn\htot{\calh = \hM \otimes \hext}
where $\hM$ describes states of the BH and $\hext$ states of the environment.  Evolution is then given by a hamiltonian $H$ which both acts within the subsystems, and acts to transfer information between the subsystems.

While ultimately incomplete, a first model of such a description is the LQFT description, based on a time-slicing of the BH geometry.  Specifically, take the slicing to be a nice slicing\LPSTU\ (for an explicit construction see \BHQIUE, eq. (3.15)).  A nice time slice crosses the horizon, and excitations on the slice either inside or outside the horizon correspond to states in $\hM$ and $\hext$ respectively.  The LQFT hamiltonian $H_{\rm LQFT}$ for a set of quantum fields then describes evolution of these excitations within $\hM$ and $\hext$, as well as transfer of excitations from $\hext$ to $\hM$, corresponding to infall of matter.  It does {\it not} describe transfer of excitations from $\hM$ to $\hext$ as this is forbidden by LQFT locality.\foot{One-way transfer of excitations by a unitary hamiltonian is permitted if the hamiltonian is not invariant under time reversal; a simple example is that of a left moving 1+1-dimensional boson.  In the present context,  the evolution of perturbations on nice slices is not T-invariant in the BH background.}  The hamiltonian also describes emergence of correlated pairs of excitations, with  excitations in $\hext$ escaping to infinity, yielding the Hawking radiation, and their partners in  $\hM$ evolving deeper into the BH.  An ultimate limitation of such a LQFT description is that both the infalling quantum matter and the Hawking radiation increase entanglement between $\hM$ and $\hext$, but that no process decreases this entanglement -- information does not escape the BH.  This appears to require BHs with arbitrarily large numbers of internal states, and, if evolution remains unitary, yields disastrous planckian BH remnants\refs{\Pres\WABHIP-\Susstrouble}.  

We expect a more complete description, if well-approximated by a subsystem description \htot, to have two new features\refs{\SGmodels,\BHQIUE}:  1) the dimension of $\hM$, or at least the number of excited BH states, shrinks as the black hole radiates, decreasing $M$; 2) to preserve unitarity, quantum information correspondingly transfers from $\hM$ to $\hext$.  The transfer of information can be precisely described in terms of transfer of entanglement\refs{\HaPr\GiShone-\Susstrans}.  

Following \refs{\NLEFTone\GiShtwo-\NVNLT}, this paper will explore the assumption that the LQFT description is a good approximate starting point for such a more complete description; that is the BH evolution, both exterior and interior, can be, for certain purposes, well approximated by the LQFT description, but that the LQFT dynamics must be corrected to properly describe unitary evolution with these two new features.  Such an approach is to be contrasted with more extreme modifications of the LQFT description\refs{\BHMR,\tHoo,\Sussholo,\Mathurrev,\Braunstein,\AMPS\PaRa,\VV}.  

Specifically, let us assume that the allowed states are in the product $\hM \otimes \hext$, and that $\hext$ is well-approximated by the space of LQFT states (Fock space or interacting generalization).  We moreover assume that the full hamiltonian is well approximated as
\eqn\hamil{H= H_{\rm int} + H_{\rm trans} + H_{\rm LQFT}\ .}
The first two terms represent dynamics beyond that of LQFT.  Specifically, in LQFT in a nice slicing, $H_{\rm LQFT}$ describes states whose evolution freezes at a definite interior radius\refs{\QBHB} due to vanishing lapse; the large number of frozen states is one manifestation of the remnant problem.  Thus, while infalling matter that has just crossed the horizon may have a good LQFT description, we expect this description to fail after sufficient time.  So, $\hM$ may, at a given time, have a factor that corresponds to LQFT states that have recently crossed the horizon (called $\hreg$ in \refs{\NLEFTone}), but does not have a complete description matching LQFT on nice slices.  Instead, we assume that there is some internal dynamics governed by $H_{\rm int}$ that acts on the states of $\hM$; {\it e.g.} it could scramble them.  

We also assume that there are interactions between $\hM$ and $\hext$ parameterized by $H_{\rm trans}$, that transfer information from inside the BH.  (As above, $H_{\rm LQFT}$ does still transfer excitations from $\hext$ to $\hM$.)
A simplest possibility is that $H_{\rm trans}$ couples excitations of $\hM$ to local operators $\calo_a(x)$ acting on $\hext$, and so is of the form
\eqn\htrans{\int dt H_{\rm trans} = -\sum_{Aa}\int dV \AA G^{Aa}(x) \calo_a(x)+ {h.c.}\ .}
Here $dV$ is the volume element for the region extending from the horizon outward, $\AA$ are operators ({\it e.g} with a unit normalization) acting on $\hM$, and $G^{Aa}(x)$ are ``transfer" coefficients parameterizing the interactions.\foot{More generally one may consider multi-local operators\refs{\NVNL}.}  Due to such interactions, the BH atmosphere ``glistens" with quantum information in a state dependent fashion.

\subsec{Characterizing information transfer}

While some of the most profound questions regard the structure of $\hM$ and its dynamics, this paper will focus on effects seen by observers near or outside the horizon.  To do so, we transform to an intermediate picture.  Let $|\Psi\rangle\in\hM\otimes\hext$ be a state evolving via \hamil, and define
\eqn\itermp{|\Psi\rangle \rightarrow U_{\rm int}(t) |\Psi\rangle\ ,}
where
\eqn\Udef{U_{\rm int}(t) = T \exp\left\{ -i \int^t dt H_{\rm int}\right\}\ .}
Then the new $|\Psi\rangle$ satisfies the Schr\"odinger equation with hamiltonian 
\eqn\Ht{H(t) = H_{\rm trans}(t) + H_{\rm LQFT}\ ,}
where now 
\eqn\htranst{\int dt H_{\rm trans}(t) = -\sum_{Aa}\int dV \AA(t) G^{Aa}(x) \calo_a(x)+ {h.c.}\ }
with
\eqn\aevol{\AA(t) = U_{\rm int}^\dagger(t) \AA U_{\rm int}(t)\ .}
Thus in the  intermediate picture explicit time dependence in $H_{\rm int}(t)$ arises from $\AA(t)$, if not already present\foot{We expect the fundamental description of the evolving black hole to be time-translation invariant; however the description of the effective dynamics in a background and with the kind of slicing we have chosen does not necessarily manifestly exhibit this invariance.  In particular, the metric in a nice slicing undergoes expansion near the horizon, which is one way of explaining the origin of the Hawking radiation.} in the coefficients $G^{Aa}(x)$.

Key questions, governed by the structure of \htranst, are whether $H_{\rm trans}$ transfers sufficient information to unitarize BH evolution, and what other effects result from such transfer.  

A simple description of transfer of information is as follows.  (For a characterization in terms of transfer of entanglement, see \refs{\HaPr\GiShone-\Susstrans}.)  Suppose we start with a collection of orthogonal states of $\hM \otimes \hext$  of the form $|i\rangle_{M} |\phi\rangle_{\rm ext}$, with $i=1,\ldots,n$; if unitary evolution by $H_{\rm trans}$ converts these to the form $|\phi\rangle_{M}|i\rangle_{\rm ext}$, where the states $|\phi\rangle$ are independent of $i$, then we say $\Delta I =\log n$ bits of information have been transferred from $\hM$ to $\hext$.  In order to do so, the interaction must convert the initial state $|\phi\rangle_{\rm ext}$ to a set of orthogonal external states.  A minimal rate of transfer of information necessary to unitarize BH evaporation is $dI/dt\sim S_{BH}/R$, where $R$ is the BH radius and $S_{BH}$ the Bekenstein-Hawking entropy, since the transfer must at least compensate for the growth of entanglement due to the Hawking process.

General interactions \htranst\ of sufficient strength clearly can transfer information at this rate, {\it e.g.} by creating excitations near the BH.  The nature of these excitations depends on
each of the terms in \htranst.  The characteristic energy dependence of $\AA(t)$ (together with possible time-dependence of $G^{Aa}$) determines the energy of outgoing excitations,\foot{Here we have chosen to define energy as measured by an asymptotic observer.}  so, {\it e.g.} transitions between widely-spaced levels of $\hM$ would emit energetic quanta.  The space dependence of $G^{Aa}$ governs both the region to which the information is transferred, and the wavelength of the excitations.  For example, one could model behavior like that of a ``firewall\AMPS" if ${G^{Aa}(x)}$ is localized within a Planck length of the horizon.  Then, even if characteristic energies are of size $\omega\sim 1/R$, \htranst\ creates excitations that an infalling observer near the horizon sees as very energetic, high-momentum particles emerging from the horizon.  Alternately, $G^{Aa}(x)$ could extend a range $\gg R$ from the black hole, describing very long-range modifications of locality.  This paper will focus on a hypothesis of nonviolent nonlocality\refs{\BHIUN,\SGmodels,\BHQIUE,\NVNL,\NLEFTone}, that the range of the interactions extends a distance $L$ from the horizon, where $L$ is neither planckian nor $\gg R$, and that the characteristic emission frequency $\omega$ is also not planckian (or microscopic, for a large black hole).  Since classical gravity already sets a scale $R$ for black hole phenomena,  a simplest assumption is that $L\sim R$ and $\omega\sim 1/R$, but conceivably other intermediate scales, {\it e.g.} $R^p$ with $p<1$, could be relevant.  The operators $\calo_a$ govern which kinds of excitations are created; these excitations carry the missing information to infinity.

A generic feature of the unitarizing  interactions \htranst\ and their generalizations is that they create {\it extra energy flux}, beyond the Hawking flux \refs{\SGmodels,\BHQIUE,\NVNL,\NLEFTone,\GiShtwo}.  This indicates\refs{\BHSM} a departure from the Bekenstein-Hawking formula for the number of black hole states, with the number of states corresponding to an entropy $S_{bh}<S_{BH}$.  While, as discussed in \BHSM, this may not lead to any fundamental inconsistency, either internally, or with established facts, it seems rather surprising.  This leads us to inquire whether there are particular forms of information-transferring evolution, {\it e.g.} as in \htranst, which do {\it not} yield extra energy flux.

If information is to escape without modifying the average Hawking energy flux, an obvious possible mechanism is for it to do so through modulation of the Hawking flux\refs{\NVNLT}.  Such modulation can be achieved if \htranst\ couples to the stress tensor, $T_{\mu\nu}$, of the BH atmosphere.  This has the added bonus of ensuring a universal mechanism for information escape, so that for example increasing the black hole evaporation rate by opening a mining\refs{\miningrefs} channel also commensurately increases the rate of escape of information, avoiding ``overfull" BHs with entropy $>S_{BH}$.  The remainder of this paper will focus on such stress-tensor couplings.

\newsec{Effects of stress tensor-induced information transfer}

\subsec{Transfer rate and effective metric fluctuations}

We thus consider evolution on $\hM\otimes \hext$ with hamiltonian of the form \Ht, where the hamiltonian $H_{\rm trans}$ couples only to the stress tensor:
\eqn\HT{\int dt H_{\rm trans}(t) = -\sum_{A}\int dV \AA(t) G^{\mu\nu}_A(x) T_{\mu\nu}(x)+ {h.c.}\ ;}
$G^{\mu\nu}_A(x)$ are a collection of tensor functions describing couplings to the BH internal states, and again $\AA$ are intermediate-picture quantum operators acting on the BH state.  Ref.~\NVNLT\ discusses conditions, in a two-dimensional example, for such couplings to not modify the average energy flux.  Here, we will focus on a different question:  if such couplings transfer sufficient information to unitarize black hole decay, what are their other effects?

The condition for sufficient information transfer was outlined above: to transfer a quantum of information, the interaction \HT\ needs to transfer an excitation of $\hM$ to one of $\hext$.  Since the focus of this paper is on the near-horizon and asymptotic physics, we are primarily interested in the effects of \HT\ on $\hext$.  To that end, we  approximate the quantum operators $\AA$ in \HT\ by classical random variables, writing
\eqn\hsource{\AA(t) G^{\mu\nu}_A(x) \rightarrow H_A^{\mu\nu}(x)\ .}
This replacement by a classical source is the {\it effective source approximation} of \refs{\NLEFTone,\GiShtwo}.
Treated as classical sources, the $H_A^{\mu\nu}$ must create sufficient excitation in $\hext$ to carry the quantum information out of the BH, beginning with, {\it e.g.}, the Unruh vacuum $|0\rangle_U$.   A benchmark for this\refs{\NVNLT}  is if the operators 
\eqn\TAS{T_A=\int dV H^{\mu\nu}_A(x) T_{\mu\nu} (x) }
are of sufficient size to create a sufficiently large collection of orthogonal states,
\eqn\orthmap{ {}_U\langle 0| T_A|0\rangle_U=0\quad ,\quad {}_U\langle 0| T_A T_B^\dagger |0\rangle_U = \delta_{AB}\ ,}
as described in the preceding section.

A related characterization, if the information is to be encoded in modifications to the Hawking radiation, with of order one quantum of information emitted per time $R$, is that the modification to the instantaneous Hawking energy flux should be of order that flux.  Notice that the coupling \HT\ behaves like a correction to the metric due to extra quantum metric fluctuations near the horizon.  The size of the corrected flux can be estimated via the effective source approximation; taking $H^{\mu\nu} = \sum_A  H^{\mu\nu}_A$, it is, to linear order in $H$, 
\eqn\pertflux{ \langle  T_{\mu\nu}(x) \rangle \simeq {}_U\vacb T_{\mu\nu}(x) \vac_U - i\int^t dV' H^{\lambda\sigma}(x')\, {}_U\vacb [T_{\mu\nu}(x),T_{\lambda\sigma}(x')] \vac_U\ .}
The equal-time commutators for the stress tensor take the general form
\eqn\stresscom{[T(x),T(x')] \sim i T \partial\delta^3(x-x') + S(x,x')\ ,}
where $S(x,x')$ is a Schwinger term\refs{\DeBo};  the first term is responsible for the tensor-transformation law of $T_{\mu\nu}$, if we perform a diffeomorphism changing the form of the metric.  The expression \stresscom\ extends to give the Green function needed in \pertflux.  For details in an explicit two-dimensional example, see \NVNLT\ [eqs. (4.7) and (4.9)].  From \pertflux, \stresscom\ we see that if $H^{\mu\nu}$ varies on spatial and temporal scales of size $R$, as described above, it will induce an $\calo(1)$ change to the stress tensor of the Hawking radiation if its amplitude is $H\sim\calo(1)$.  

Our conclusion therefore is:  {\it For interactions \HT\ to transfer information to the outgoing Hawking radiation at the necessary rate $\sim 1/R$, the couplings $G^{\mu\nu}_A$ should have size of order unity at frequencies and wavenumbers of size $\sim 1/R$, that is the corresponding effective source $H^{\mu\nu}(x)$ is a function with magnitude $\calo(1)$, with characteristic frequencies and wavenumbers $\sim 1/R$.}\foot{Note that this argument gives the necessary strength of interactions to unitarize black hole disintegration, if by this mechanism.  It does not, yet, give a description of the complete unitarized  amplitudes.}  To ensure regularity at the horizon, these functions can be specified in terms of coordinates that are regular there,  {\it e.g.} Kruskal coordinates $(U,V,\theta,\phi)$ or 
Eddington-Finkelstein coordinates $(v,r,\theta,\phi)$ or,  for Kerr black holes, generalization of the latter to 
Kerr coordinates\refs\MTW.

If $H^{\mu\nu}$ is thought of as a correction to the metric, we have just found that this metric needs to be {\it strongly fluctuating} in the region just outside the horizon for \HT\ to describe sufficient escape of information.  Note that while the effective metric is strongly fluctuating,  the corresponding fluctuations are relatively soft, if their characteristic momentum scales are $\sim 1/R$.  For example, typical tidal forces they induce are of the same size as due to the the background curvature at the horizon, $\sim 1/R^2$.  Of course such fluctuations also exhibit the potential for yielding singularities in the effective metric, raising the question of regularity criteria from the underlying framework.  For regular fluctuating metrics ({\it e.g.} with moderate curvature), we find the interesting possibility that the metric could be strongly quantum, but in a manner that doesn't directly involve Planck scale physics -- and that can be nonviolent to infalling observers.  The fluctuations may be strong but soft.

\subsec{Propagation in fluctuating BH atmospheres}

If the effective description of BH dynamics is via such a strongly fluctuating effective metric, with $\calo(1)$ fluctuations in a region of size $\calo(R)$ surrounding the horizon, the coupling \HT\ will imply nontrivial effects on quantum fields in the vicinity of the horizon.  Specifically, \HT\ will have an effect both on infalling matter, and on emitted light from the zone near the horizon.  

To estimate these effects, we work in the approximation where particles (infalling matter, outgoing light) follow geodesics; with the additional coupling
\eqn\Hcoup{\int dt H_{\rm trans}(t) \rightarrow -{\epsilon\over 2} \int dV H^{\mu\nu}(x) T_{\mu\nu}(x)\ ,}
these are geodesics in the perturbed metric
\eqn\metpert{{\tilde g}^{\mu\nu} = g^{\mu\nu}+ \epsilon H^{\mu\nu}\ ,}
where $g^{\mu\nu}$ is the classical metric, {\it e.g.} Schwarzschild or Kerr. Although the perturbation has been argued to be $\calo(1)$ we will introduce a parameter $\epsilon$ in order to parameterize the couplings and estimate the size of their effects.
The geodesic equation can be written in terms of the time variable of the original metric, but that will no longer be an affine parameter for the new metric, resulting in additional terms:
\eqn\newgeod{{d^2\tX^\mu\over dt^2} = -{\tilde \Gamma}^\mu_{\nu\lambda} {d\tX^\nu\over dt} {d\tX^\lambda\over dt} +   {\tilde \Gamma}^0_{\nu\lambda} {d\tX^\nu\over dt} {d\tX^\lambda\over dt} {d\tX^\mu\over dt} \ .}
Solutions may be written in terms of deviation from the solution for the unperturbed metric, ${\tilde X}^\mu = X^\mu + \delta X^\mu$.  
The equation for the geodesic perturbation is
\eqn\geodp{\eqalign{ {d^2\delta X^\mu\over dt^2} &= -\delta {\Gamma}^\mu_{\nu\lambda} {dX^\nu\over dt} {dX^\lambda\over dt} +  \delta  \Gamma^0_{\nu\lambda} {dX^\nu\over dt} {dX^\lambda\over dt} {dX^\mu\over dt}\cr &- 2 {\Gamma}^\mu_{\nu\lambda} {d\delta X^\nu\over dt} {dX^\lambda\over dt} +  \Gamma^0_{\nu\lambda}{dX^\nu\over dt} \left( 2 {d\delta X^\lambda\over dt} {dX^\mu\over dt} + {dX^\lambda\over dt} {d\delta X^\mu\over dt}\right)\ ,}}
where
\eqn\chpert{\delta \Gamma^\mu_{\nu\lambda} = {\epsilon \over 2}\left( g_{\nu\rho} g_{\lambda\sigma} \nabla^\mu H^{\rho\sigma} - g_{\nu\rho} \nabla_\lambda H^{\mu\rho} - g_{\lambda\sigma} \nabla _\nu H^{\mu\sigma}\right)\ .}
So, if $H\sim \calo(1)$ and varies on a spacetime scale $L$, the terms on the first line of \geodp\ will drive order $\epsilon$ changes to $d{\tilde X}^i/ dt$ once  integrated over times $t\sim L$ -- hence $\calo(1)$ changes for $\epsilon=1$.

\subsec{Possible observational windows}

Concretely, suppose that quantum effects lead to fluctuations in the effective metric, coupled as in \Hcoup,  with size $\epsilon H\sim \calo(\epsilon)$ and spacetime variation on a scale $L$, in a region extending to a radius $R_a$ outside the horizon of an astrophysical black hole. As noted above, the simplest case is $L\sim R$, and likewise $R_a\sim  R$ is a simplest assumption.  Such fluctuations could have observational implications via at least four types of observable phenomena.

\subsubsec{Accretion disk dynamics}

The classical orbits assumed in models of accretion-disk dynamics will be significantly perturbed in the region $r<R_a$.  This presents a source of disruption of the accretion flow in the near-horizon atmosphere region.\foot{I thank C. Reynolds for suggesting the possible relevance of such an effect, and O. Blaes for explanations of current observational constraints.}  One can readily estimate the magnitude of the effect, under the preceding assumptions:  by the time a free particle moves through a distance $\sim L$, it will experience an order $\epsilon$ deflection of its velocity, $\delta dX^i/dt \sim \epsilon$.  Specifically, with $L\sim R$ and $\epsilon=1$  this means significant perturbation after one orbit -- for order one metric perturbations, regular free orbits are substantially deformed in the region $r<R_a$.  Current observations may already constrain such effects.  For example, X-ray binaries in the high/soft state exhibit a near-thermal spectrum with its hard end attributed to orbits approaching the innermost stable circular orbit, which for a Schwarzschild BH is at $r=3R$. While calculation of properties of the accretion disk in this regime involves complicated collisional magnetohydrodynamics, one expects metric fluctuations of sufficient strength and range to perturb these flows, possibly providing a source of turbulence and/or increasing infall rates. 

\subsubsec{Imaged features from accreting matter}

Light emitted from an accreting classical BH  has distinctive features, such as a black hole shadow\refs{\Bardeen,\Falcke} and photon ring\refs{\JoPs}\foot{For discussion of testing deviations from Kerr by observing these features, see \refs{\BaFr,\BaYo,\JoPs}}.  These 
arise when photons emitted from material in the near-horizon accretion disk are strongly lensed by the near-horizon geometry; for example the photon ring results from null geodesics that orbit the black hole multiple times near the photon orbit.  
 In addition to the possible distortion of the matter orbits, strong metric fluctuations will produce additional lensing contributions, which will fluctuate with the effective metric perturbation: light scintillates in the BH atmosphere.  For light emitted from the near-horizon region and traveling a distance $\sim L$, one estimates a typical angular deflection $\delta \theta \sim \epsilon$, so again order unity for $\epsilon=1$.  
Thus, large metric fluctuations have the potential to significantly alter the expected electromagnetic (sub-mm) features, {\it e.g.} distorting the edge of the shadow and suppressing the photon ring.  The latter is a particularly attractive target, since it is sensitive to the geometry in the vicinity of what would be the photon orbit, which in the small angular momentum limit lies quite close to the horizon, at $r=3R/2$.  Distortion of the effective metric in this region could produce the signature of a distorted or missing photon ring.

In particular, the Event Horizon Telescope\foot{For discussion of the instrument see \refs{\Fishetal} and for capabilities see \refs{\BJLP}.} is planned to achieve sensitivity capable of resolving Sagittarius A* at an angular resolution of approaching 10 $\mu as$ through mm and sub-mm wavelength observations in the near term, and thus is expected to be capable of resolving the shadow and photon ring characteristic of its anticipated near-horizon Kerr geometry.  According to the above discussion, the quantum effects we have described are therefore in the range of projected experimental sensitivity.  This is a potentially important experimental opportunity to explore, and observation of any such effects could have truly profound implications.  

\subsubsec{Lensing distant objects}

While opportunities may be rare to observe a distant object pass behind a BH such that lensing due to the gravitational field of the BH's atmosphere is probed, this in principle offers another observational window toward probing BH scintillation due to fluctuations in the effective metric.  Specifically, lensing where a significant part of the light rays cross through the region $r<R_a$ will experience, as with the above, angular distortions of size $\delta \theta \sim \epsilon$.  Lensing opportunities may also arise in binaries where one member is a black hole.

\subsubsec{Gravitational inspiral}

Another possible window is that of gravitational radiation emitted during gravitational inspiral,\foot{For similar comments on possible modifications to gravitational-wave signatures, see \AMPSS.} {\it e.g.} as a BH captures a neutron star, a process expected to be observable by Advanced LIGO.  This process probes the near-horizon and strong-field regime.  Perturbations of the neutron star orbit near the horizon are expected to distort gravitational waveforms arising from emission during the final stages of inspiral.

For each of these possible effects, one should state the important caveat that a sharp prediction for the timescale at which the new information-transferring effects become important is still missing.  There are constraints that the new effects should not begin for black holes younger than a time $\sim R \log (R/l_{\rm Planck})$ but must begin by a time $\sim R(R/l_{\rm Planck})^2$; for an astrophysical black hole this represents a large range of timescales.  Specifically, for Sgr A*, the corresponding range of timescales is from about 8 hours to $10^{74}$ times the present age of the Universe.

\subsubsec{Simulation}

Despite the relative simplicity of eqs. \geodp, \chpert, their analytic solution for specific perturbations is typically problematic, and moreover we do not have a specific prediction for the form of these perturbations, which may have a rather random apparent character.  Also, treatment of accretion disk dynamics, accretion disk images, and gravitational inspiral is typically performed through numerical simulation.  This indicates a numerical approach to investigating effects of black hole glistening.  Specifically, in a numerical approach the geodesic equation is augmented by the perturbation terms \geodp.  To describe the metric perturbations entering \chpert, regularity is most easily enforced by working in regular coordinates at the horizon, {\it e.g.} Kerr coordinates\MTW\ $(v,r,\theta,\tilde\phi)$ for the Kerr solution.  Then, the effects may be modeled by considering a superposition of random perturbations to Kerr metric components in these coordinates, which for example take the form
\eqn\kerrpert{H^{\mu\nu} \sim f(r,v) e^{-i\omega v+ ik r} Y_{lm}(\theta,\tilde\phi)}
where $f$ is an $\calo(1)$ smooth window function localized to an order $R_a$-sized range of $r$ outside the horizon, and to a temporal range $\Delta v$, the frequencies $\omega$, $k$ are of size $1/L$, and, {\it e.g.}, $l,m$ take some moderate values.  As noted above, the simplest case to explore is where all scales are equal:  $R_a\sim \Delta v\sim L\sim R$.  One may equally well choose to consider random Christoffel corrections \chpert\ of the corresponding size.

\newsec{Concluding comments}

Typically, when one envisions strong quantum gravitational fluctuations, one imagines strong fluctuations associated with Planck-sized curvatures.  This paper has explored a different possibility, that the metric could be strongly fluctuating, but on spacetime distance scales that are much larger than the Planck length.  If so, observers may experience important effects\refs{\SGmodels,\BHQIUE,\NVNL}, but not necessarily high-energy or hard effects\AMPS; while the geometry is not semiclassical, it also can be nonviolent.  This of course assumes that such fluctuations, while strong, respect appropriate regularity conditions on the metric.   

The proposal that such strong but soft fluctuations could play a role in near-horizon black hole dynamics is not  ad hoc.  To unitarize black hole evaporation, {\it some} new effects must transfer information out of a black hole, or even stranger modifications to physics are needed.  If this transfer occurs while preserving, at least in an an approximation, a spacetime picture of the near-horizon geometry, the transfer is most simply described in terms of new couplings to operators acting on quantum fields near the horizon.  While ultimately we expect to require a more fundamental description of the dynamics (for some thoughts on this, see \BHQIUE), this effective approach of considering modifications of local quantum field theory, if valid in an approximation, is anticipated to supply important guidance.  

There are significant constraints on such an effective description.   Transfer that couples through the stress tensor\refs{\NVNL,\NVNLT}, as in \HT, has some advantages:  1) it offers the possibility of minimal extra energy flux from the BH, beyond that of Hawking\refs{\NVNLT}, improving the prospects that the Bekenstein-Hawking entropy approximately gives the BH density of states\refs{\BHSM}; 2) it is a universal coupling, rather than to specific fields, as might be anticipated for a new quantum-gravitational effect; 3) the universality of such a coupling also helps provide a consistent way to address various Gedanken experiments, for example involving the possibility of mining energy from a BH\refs{\NLEFTone\GiShtwo,\NVNLT}.  

If the information necessary to unitarize evaporation does transfer through such stress tensor couplings, the necessary rate of transfer determines the magnitude of these couplings.  These couplings can moreover be effectively described in terms of  quantum fluctuations of the metric, above and beyond fluctuations one would expect in the usual QFT quantization near the BH horizon.  The necessary rate indicates that these fluctuations should lead to contributions  to the effective metric approaching $\calo(1)$ in magnitude, and so these are strong in the sense just described.

Let us also compare the picture described in this paper to other proposed scenarios for unitary BH evolution.  

Firewalls\AMPS\ are envisioned to arise from transfer of information to within a microscopic ({\it e.g.} Planck) distance of the horizon.  As we have noted, they can possibly be modeled by interactions of the form \htranst, but the ``sharpness" of the information transfer implies that this transfer is accompanied by production of energetic particles at the horizon.  On the other hand, the firewall scenario assumes that local QFT holds everywhere just outside the horizon, indicating that the new physics cannot be probed unless one is sensitive to dynamics immediately {\it at} the horizon.  

The fuzzball scenario\refs{\Mathurrev} does not appear to make a clear statement about how far modifications to classical geometry extend from the horizon.  However, these modifications involve stringy higher-dimensional geometry, and are expected to be hard (high-momentum), and for example to be characterized by large average squared curvatures. While there are conjectures\refs{\MaPl} that such hard modifications do not affect infalling matter and observers, through a new version of complementarity, these remain controversial.  If fuzzball modifications do extend sufficiently far outside the would-be horizon, and can be quantified for Kerr or Schwarzschild black holes, they could likewise affect near-horizon observations, and these modifications might be quantified and simulated as described above. Such modifications may of course have even more radical effects than described herein.  In fact fuzzballs are an example of the general scenario of massive remnants\refs{\BHMR}, where the near-horizon geometry is replaced by a new kind of physical interface.  Such a departure from the BH geometry might be observed through the phenomena described in the preceding section, but also is expected  yield even more pronounced signatures through impact of material on the  structure of the interface\refs{\BLR,\SBGlett}.

The possible importance of significant fluctuations outside the horizon described herein bears some rough similarity to a scenario proposed in \refs{\DvaliAA}; there, a black hole is described as a collection of $\sim S_{\rm BH}$ gravitons with wavelength $\sim R$.  In such a picture one might also expect the geometry to have significant quantum fluctuations on scales $\sim R$ and outside the would-be horizon, and expect that those fluctuations are soft.  In the present paper it is anticipated that at least some of the features of the semiclassical geometry remain intact.  Important challenges for the picture of \DvaliAA\ include those of giving a correct description of the nonlinear interactions between gravitons and of making contact with the description of the semiclassical geometry; for recent ideas regarding the latter, see \DvaliEJA.

In conclusion, if gravity respects the principles of quantum mechanics, this indicates the need for new quantum phenomena extending to and beyond the horizon of a black hole, even when the black hole is large.  A simple possibility is new interactions, transferring information.  Multiple considerations suggest that these could proceed through couplings to the stress tensor of fields in the black hole atmosphere, behaving like  quantum fluctuations of the metric; necessary rates of information transfer indicate that these fluctuations would be significant, though they could remain ``soft."  And, if such effects extend over a range of order the BH radius $R$, this implies the possibility of observing them through consequences of perturbating the near-horizon classical metric, 
including possible disruption of near-horizon accretion disk dynamics and gravitational lensing.  Existing observations of accretion-disk phenomena may provide important constraints on such new effects, and near-term observations are expected to open up new windows of possible sensitivity, beginning with the Event Horizon Telescope and detection of gravity waves from inspirals.

\bigskip\bigskip\centerline{{\bf Acknowledgments}}\nobreak

I thank O. Blaes, J. Hartle, and D. Marolf for helpful conversations.  
I would especially like to thank S. Britzen for providing useful background on the EHT and capabilities, for valuable conversations, and for inviting me to attend the COST conference ``99 years of Black Holes -- from Astronomy to Quantum Gravity,"  where a preliminary version of this discussion was presented.  Questions from and conversations with participants at this conference were valuable, particularly those with C. Reynolds and V. Fish. 
This work  was supported in part by the Department of Energy under Contract DE-FG02-91ER40618 and by  Foundational Questions Institute grant number FQXi-RFP3-1330.

\listrefs
\end